\documentclass[preprint,tightenlines,aps,prc,showpacs,nofootinbib,showkeys]{revtex4}
\usepackage{graphicx,amsmath,amssymb}
\usepackage{amssymb,latexsym,revsymb,comment}
\usepackage{epsfig,bm}
\newcommand{\be}{\begin{equation}}
\newcommand{\ee}{\end{equation}}
\newcommand{\bea}{\begin{eqnarray}}
\newcommand{\eea}{\end{eqnarray}}

\begin{document}

\sloppy

\def\a{{\alpha}}
\def\b{{\beta}}
\def\d{{\delta}}
\def\D{{\Delta}}
\def\e{{\varepsilon}}
\def\g{{\gamma}}
\def\G{{\Gamma}}
\def\k{{\kappa}}
\def\l{{\lambda}}
\def\L{{\Lambda}}
\def\m{{\mu}}
\def\n{{\nu}}
\def\bfk{{\bm k}}
\def\o{{\omega}}
\def\O{{\Omega}}
\def\S{{\Sigma}}
\def\s{{\sigma}}
\def\th{{\theta}}
\def\x{{\xi}}
\def\cO{\mathcal{O}}
\def\eps{{\varepsilon}}
\def\epsp{{\varepsilon^\prime}}
\def\epspp{{\varepsilon^{\prime\prime}}}
\def\lp{{\lambda^\prime}}
\def\kppperp{{\bm{k}^{\prime \prime \perp}}}
\def\gp{{\gamma^+}}
\def\lamp{{\lambda^\prime}}
\def\kpr{{\mathbf{k}^\perp_{\text{rel}}}}
\def\g5{{\gamma^5}}
\def\kpperp{{\bm{k}^{\prime \perp}}}
\def\xt{{\tilde{x}}}
\def\kpp{{\bm{k}^{\prime\perp}}}
\def\kppp{{\mb{k}^{\prime\prime\perp}}}
\def\ppp{{\mb{p}^{\prime\perp}}}
\def\dperp{{\bm{\Delta}^\perp}}
\def\koneperp{{\bm{k}^{\perp}_{1}}}
\def\ktwoperp{{\bm{k}^{\perp}_{2}}}
\def\pperp{{\bm{p}^{\perp}}}
\def\Pperp{{\bm{P}^{\perp}}}
\def\koneplus{{k^{+}_{1}}}
\def\ktwoplus{{k^{+}_{2}}}
\def\Pplus{{P^{+}}}
\def\pplus{{p^{+}}}
\def\qperp{{\bm{q}^{\perp}}}
\def\pmag{{p^{\perp}}}
\def\kperp{{\bm{k}^{\perp}}}
\def\kplus{{k^{+}}}
\def\kaperp{{\bm{\kappa}^{\perp}}}
\def\kaplus{{\kappa^{+}}}
\def\Pee{{P^{\mu}}}
\def\Pp{{{P}^{\prime}{}^{\mu}}}
\def\Pbar{{\bar{P}^{\mu}}}
\def\Pbarplus{{\bar{P}^{+}}} 
\def\ym{{y^{-}}}
\def\Ppbra{{\langle \; P^{\prime} \; |}}
\def\Pket{{ | \; P \; \rangle}}
\def\phid{{\hat{\phi}^\dagger}}
\def\phiud{{\hat{\phi}}}
\def\dplus{{\partial^{+}}}
\def\meask{{\frac{d\kplus  d\kperp}{\sqrt{ 2 \kplus} (2\pi)^3}}}
\def\measpj{{\frac{d\pplus_{j}  d\pperp_{j}}{\sqrt{ 2 p^+_{j}} (2\pi)^3}}}
\def\measka{{\frac{d\kaplus  d\kaperp}{\sqrt{ 2 \kaplus} (2\pi)^3}}}
\def\ak{{\hat{a}(\kplus,\kperp)}}
\def\adk{{\hat{a}^\dagger(\kplus,\kperp)}}
\def\aka{{\hat{a}(\kaplus,\kaperp)}}
\def\adka{{\hat{a}^\dagger(\kaplus,\kaperp)}}
\def\pket{{ | \; p_{1}, p_{2} \; \rangle}}
\def\qbra{{\langle \; p_{3}, p_{4}  \; |}}
\def\ndpd{{\mathcal{F}(X, \zeta, t)}}
\def\ie{{i \epsilon}}
\def\ofpd{{F(x,\xi,t)}}
\def\dw{{D_{\text{W}}}}
\def\bigdw{{\Delta_{\text{W}}}}
\def\mn{{\langle \mu^2 \rangle}}
\def\pp{{p^{\prime}}}
\def\lam{{\Lambda}}
\def\kminus{{k^{-}}}
\def\pminus{{p^{-}}}
\def\qminus{{q^{-}}}
\def\qplus{{q^+}}
\def\Gt{{\tilde{G}}}
\def\pe{{\mathcal{P}}}
\def\qu{{\mathcal{Q}}}
\def\psiket{{|\psi\rangle}}
\def\psitwoket{{|\psi_{2}\rangle}}
\def\psiquket{{|\psi_{\qu}\rangle}}
\def\psibra{{\langle\psi|}}
\def\psitwobra{{\langle\psi_{2}|}}
\def\psiqubra{{\langle\psi_{\qu}|}}
\def\xpp{{x^{\prime\prime}}}
\def\yp{{y^\prime}}
\def\xp{{x^\prime}}
\def\zp{{z^\prime}}
\def\zpp{{z^{\prime\prime}}}
\def\ypp{{y^{\prime\prime}}}
\def\bfQ{{\bm Q}}

\newcommand{\eq}[1]{Eq.~(\ref{#1})}
\def\ol#1{{\overline{#1}}}

\preprint{NT@UW-09-24}
\preprint{UMD-40762-470}
\preprint{INT-PUB-09-058}

\title{The Relation Between Equal-Time and Light-Front Wave Functions}

\author{Gerald A.~Miller}
\email[]{miller@phys.washington.edu}
\affiliation{%
Department of Physics\\ 
University of Washington\\ 
Seattle, WA 98195-1560
USA
}

\author{Brian~C.~Tiburzi}
\email[]{bctiburz@umd.edu}
\affiliation{%
Maryland Center for Fundamental Physics\\
Department of Physics, 
University of Maryland\\
College Park,  
MD 20742-4111, 
USA
}

\begin{abstract}

The relation between equal-time and light-front wave functions is
studied using models for which the four-dimensional solution of the Bethe-Salpeter wave function can be obtained. 
The popular prescription of defining the longitudinal momentum fraction using the instant-form free kinetic energy and third component of momentum 
is found to be incorrect except in the non-relativistic limit. 
The only presently known way to obtain light-front wave functions from rest-frame,  
instant-form wave functions is to boost the latter wave functions to the infinite momentum frame.
Despite this fact, 
we prove a relation between certain integrals of the equal-time and light-front wave functions.

\end{abstract}

\maketitle

\section{Introduction} \label{chap:intro}

Light-front hadronic wave functions are used to interpret a variety of high energy hadronic processes  
and experimentally observable quantities including:
electromagnetic form factors~%
\cite{Drell:1969km,West:1970av,Lepage:1979zb,Jacob:1985bc}, 
estimates of weak decay rates~%
\cite{Brodsky:1998hn,Choi:2007yu},  
quark recombination in heavy ion collisions~%
\cite{Fries:2003vb,Fries:2003kq,Hong:2005rp}, 
coherent pion production of di-jets~%
\cite{Frankfurt:1993it,Frankfurt:1999tq,Frankfurt:2000jm},
single spin asymmetries in semi-inclusive deep inelastic scattering~%
\cite{Brodsky:2002cx,Gamberg:2007wm},
computing various high-energy scattering amplitudes using the color dipole approach~%
\cite{Kopeliovich:1981pz,Mueller:1989st,Kopeliovich:2009cx,Kopeliovich:2009yw}, 
computing the cross sections for electromagnetic production of vector mesons 
\cite{Radyushkin:1996ru,Vanderhaeghen:1999xj,Goloskokov:2006hr},
and heavy quark fragmentation in the quark gluon plasma~%
\cite{Sharma:2009hn}.
Therefore it is useful to understand how to obtain light-front wave functions from a fundamental point of view.

There is a large body of knowledge regarding techniques, models and insights
related to the equal-time rest-frame (ETRF) formalism. 
For example, 
spectroscopy is typically handled using this formalism.  
It is therefore natural to try to relate the ETRF  
wave function with the light-front wave function. 
One popular method uses a recipe to convert the spatial momenta of the constituents,
$\bm{k}_i$, 
into light-front momenta,
$( x_i, \bm{k}_{i \perp} )$.
To be concrete,   
consider a bound state composed of two equal-mass constituents without spin. 
In this case, 
the ETRF wave function depends on the momentum 
$\bm{k}$ 
of one constituent. 
The recipe to convert the ETRF wave function to a light-front wave function
is to introduce the longitudinal momentum fraction by the relation
\begin{equation}  \label{eqn:bad}
x 
= 
\frac{k^+}{P^+} 
=
\frac{E_{\bm{k}}+k^3}{2E_{\bm{k}}}
= 
\frac{1}{2} + \frac{k^3}{2\sqrt{\bm{k}_\perp^2 + (k^3)^2 + m^2}}
,\end{equation}
where the single-particle energy is given by 
\be
E_{\bm{k}}=\sqrt{\bfk^2+m^2}
,\ee
and 
$P^+$ 
is the plus-component of the total momentum,
$P$, 
of the bound state.%
\footnote{
For any Lorentz four-vector 
$A^\mu$,
we define light-cone coordinates, 
$A^\pm$, 
by 
$A^\pm = A^0 \pm A^3$. 
Readers who employ a factor of 
$1 / \sqrt{2}$ 
to define their light-cone coordinates should note that only one equation in this work
depends on the choice of convention.
This equation is an intermediate step appearing in Eq.~\eqref{convention}.
}
Using the recipe in \eq{eqn:bad} on a function of the single-particle energy invokes the change of variables
\be
f ( \bfk^2+m^2) 
\longrightarrow
f \left( \frac{\bfk_\perp^2+m^2}{4 x(1-x)} \right) 
\notag
.\ee 
The latter form looks like the argument of a light-front wave function. 
The recipe to construct a light-front wave function from an ETRF wave function often also includes a Jacobian factor, 
$\sqrt{J} = \sqrt{ \partial k^3 / \partial x}$,
to preserve the wave function normalization.

The relation in Eq.~\eqref{eqn:bad}, 
however, 
appears to neglect any binding effect. 
While it is true in general that the plus momentum is additive~\cite{Brodsky:1997de}, 
$P^+ = \sum_i k^+_i$, 
the energy of the bound state is not, 
$P^0 \neq \sum_i E_{\bm{k}_i}$.  
This leads one to suspect that there is nothing fundamental about 
making light-front wave functions by following the popular recipe. 
In fact, 
the issue can be resolved, 
because the formal relationship between the ETRF and the light-front
wave functions has been known for a long time. 
Both involve energy integrals of the four-dimensional Bethe-Salpeter wave function, 
$\Psi(k,P)$: 
over 
$k^0$ 
in the case of the ETRF, 
and 
over
$k^-$ 
in the case of the light-front formulation. 
Given the covariant wave function
$\Psi$, 
one can study the relationship between the ETRF and light-front wave functions. 
The purpose of this paper is to provide such a study for a set of simple models. 
Although the treatment of particles with spin can be handled after suitable 
regularization~\cite{Tiburzi:2002tq,Tiburzi:2004mh}, 
we consider only spin-zero systems made of two spinless constituents
of equal mass throughout to simplify the presentation.

Here is an outline of our approach, and summary of our findings. 
Sect.~II is concerned with two-body bound states in covariant field theory and the Bethe-Salpeter equation. 
In particular, 
the explicit relation between the light-front 
($LF$) 
and rest-frame instant-form wave functions 
($IF$) 
and the solution of the Bethe-Salpeter equation is discussed.  
Next, 
in Sect.~III an exactly soluble model  involving point-like coupling of a hadron to two scalar constituents is introduced to compare the  light-cone 
and familiar instant-form wave functions.  
We find the simple transformation in Eq.~\eqref{eqn:bad} does not relate the $IF$ wave function to the $LF$ wave function, 
except in the non-relativistic limit.
Further it is verified that boosting the ETRF wave function to infinite momentum produces the light-front wave function.
Sect.~IV investigates solutions of the Bethe-Salpeter wave function by means of the Nakanishi integral representation.
Similarly we find that the $IF$ wave function is not related to the $LF$ wave function by Eq.~\eqref{eqn:bad}. 
For the general class of models of the Nakanishi type, 
we are able to show that the ETRF and light-front wave function agree in the non-relativistic limit, 
and that boosting the ETRF wave function to infinite momentum produces the light-front wave function. 
In Sect.~V, 
we summarize our work, 
and show that, 
despite the failure of the recipe to relate $IF$ and $LF$ wave functions,
certain integrals of these wave functions are identical.

\section{Bethe-Salpeter Equation and Bound states} \label{bound}              %

\begin{figure}
\begin{center}
\epsfig{file=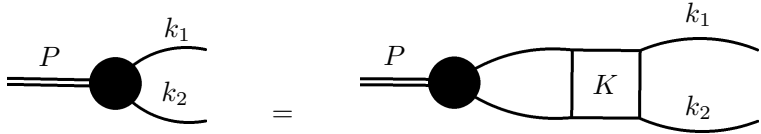,width=10cm}
\end{center}
\caption{Diagrammatic representation of the Bethe-Salpeter equation. 
The blob represents the vertex function 
$\Gamma$, 
and the total momentum is 
$P$.
}
\label{fBS}
\end{figure}

We first discuss two-body bound states in covariant field theory. 
In terms of fully covariant operators, 
the Lippmann-Schwinger equation for the two-particle transition matrix 
$T$ 
appears as
\begin{equation} \label{LS}
T = K + K G T.
\end{equation}
Above, 
$K$ 
is the irreducible two-particle scattering kernel
and 
$G$ 
is the completely disconnected two-particle propagator, 
which is merely the product of two single-particle propagators. 
A pole in the 
$T$-matrix 
(at some value of the total momentum-squared, 
$P^2 = M^2$, 
say) 
corresponds to a two-particle bound state of mass 
$M$. 
Investigation of the pole's residue gives an equation for the bound state vertex 
$\Gamma$
\begin{equation} \label{eqn:Gamma}
\Gamma = K G \Gamma,
\end{equation}
see Fig.~\ref{fBS}. 
The bound-state amplitude 
$\Phi$ 
is defined as 
$G \Gamma$ 
and hence satisfies a similar equation, 
the Bethe-Salpeter equation (BSE)~%
\cite{Schwinger:1951ex,GellMann:1951rw,Salpeter:1951sz}
\begin{equation} \label{eqn:BetheSalpeter}
\Phi = G K \Phi
.\end{equation}

In the momentum representation and using the notation of~%
\cite{Carbonell:2008tz}, 
the BSE for two spinless particles reads:
\begin{eqnarray}
\label{bs}
\Phi(k, P)
&=&
G \Big( k + \frac{P}{2}, k - \frac{P}{2} \Big)
\int \frac{d^4k'}{(2\pi)^4} iK(k,k',P)\Phi(k', P)
.\end{eqnarray}
The total momentum of the bound-state is 
$P$, 
while the momenta of the constituents are
$k_1 = k + \frac{1}{2} P$, 
and 
$k_2 = k - \frac{1}{2} P$. 
The relative momentum of the two constituents is then
$k = \frac{1}{2} (k_1 - k_2)$. 
This form makes manifest the symmetry between the two particles. 
We also find it convenient to utilize a form of the BSE that is asymmetric. 
In this alternate form, 
we denote the bound-state amplitude by 
$\Psi(k_1, P)$, 
where 
$k_1$
is the momentum of one of the particles. 
The relation between the two amplitudes is
\be
\Psi(k_1, P)
=
\Phi \Big( k_1 - \frac{P}{2}, P \Big)
.\ee
We will often treat the subscript as implicit.

Armed with the Bethe-Salpeter amplitude 
$\Psi (k_1, P)$,
one can calculate field-theoretic bound-state matrix elements by taking the appropriate residues of four-point Green's functions. 
These matrix elements may ultimately require knowledge of higher-point functions which then must be solved for consistently in the same dynamics. 
The Bethe-Salpeter amplitude 
$\Psi (k_1, P)$
is in some ways the covariant analogue of the Schr\"odinger wavefunction. 
While the features of relativistic field theory (in particular: particle creation and annihilation, retardation effects, $\ldots$) 
make the exact analogy impossible, 
in the non-relativistic limit, 
one can show that the BSE reduces to the Schr\"odinger equation.

The above discussion contains a graphical derivation of the BSE. 
It is useful to recall the field-theoretic coordinate-space definition of the Bethe-Salpeter wave function
\bea
\Psi(x_1,x_2,P)
=
\langle0|T \{ \phi(x_1)\phi(x_2) \} |P \rangle
,\label{ftpsi}
\eea
where the constituent fields are denoted by 
$\phi$.
One obtains the relation with 
$\Psi(k_1,P)$ 
by appealing to space-time translational invariance 
\bea
\Psi(x_1,x_2,P)
=
\Psi'(x_1-x_2,P)
\,
\exp[-iP\cdot(x_1+x_2)/2] 
,\label{coord}
\eea
and realizing that the Fourier transform is the amplitude 
$\Phi(k,P)$ 
above,
namely
\bea
\Phi(k, P)
=
\int  d^4z \, \Psi' (z,P) \exp (ik\cdot z )
.\label{psihat}\eea
Projecting the constituents onto states of definite four-momentum, 
we indeed find
\be
\int d^4 x_1 d^4 x_2 \,
\Psi(x_1,x_2,P)
\exp(ik_1\cdot x_1+ik_2 \cdot x_2)
= 
(2 \pi)^4 \delta^{(4)} (P - k_1 - k_2) 
\, \Psi(k_1, P)
.\ee

The relation between three-dimensional wave functions and the Bethe-Salpeter wave function 
emerges from restricting the latter function to the corresponding initial boundary. 
In the case of light-front dynamics, 
the boundary surface is customarily defined on the plane 
$x^+ =0$; 
while, 
for instant-form dynamics, 
the boundary surface is specified by the origin of time,
$x^0 = 0$.
To carry out the projection onto the light front, 
one starts from an integral 
$I(k_1, k_2, P)$ 
that restricts the variation of the arguments of the latter function to the light-front plane. 
This plane is generally defined by the condition 
$\omega\cdot x=0$, 
where 
$\omega$ 
is an arbitary four-vector with 
$\omega^2=0$~%
\cite{Carbonell:1998rj}. 
The light-front integral 
$I(k_1,k_2,P)$ 
is defined by the equation:
\bea 
I(k_1,k_2,P)
\equiv 
\int d^4x_1d^4x_2 \, \delta(x^+_1)\delta( x^+_2)
\Psi(x_1,x_2,P)
\exp(ik_1\cdot x_1+ik_2 \cdot x_2)
.\label{idef}
\eea
This integral does not produce the covariant momentum-space Bethe-Salpeter amplitude, 
rather the projection
\be
I(k_1,k_2,P)
=
(2 \pi)^3 \delta^{(+,\perp)} ( P - k_1 - k_2)
\int _{-\infty}^\infty \frac{dk_1^-}{ 2 \pi}
\Psi(k_1,P)
\label{firsti}
.\ee
The delta-function appearing above is three-dimensional, 
$\delta^{(+,\perp)} (k) \equiv \delta(k^+) \delta( \bm{k}_\perp)$.

We can obtain another expression for 
$I(k_1, k_2, P)$
involving the light-front wave function,
and thereby deduce the relation with the covariant wave function. 
The valence light-front wave function is the coefficient of the valence state in the Fock-space expansion of 
$| P \rangle  \equiv | P^+, \bm{P}_\perp \rangle $.  
On the light-front, 
the bound state
$| P \rangle$
is chosen to satisfy the covariant normalization condition, 
$\langle P' | P \rangle = 2 P^+ (2\pi)^3 \delta^{(+,\perp)} (P' - P)$, 
and has the light-front Fock space expansion
\bea
| P \rangle
=
\frac{1}{\sqrt{2 \mathcal{Q}}}
\int 
\frac{dk_1^+ d \bm{k}_{1\perp}}{2 k_1^+ (2\pi)^3}  
\frac{dk_2^+ d \bm{k}_{2\perp}}{2 k_2^+ (2\pi)^3} 
\;
\psi_{LF} (k_1, k_2, P) 
\;
2 P^+ (2\pi)^3 \delta^{(+,\perp)} ( P - k_1 - k_2)
\;
a_{k_1}^\dagger a_{k_2}^\dagger | 0 \rangle
.
\notag \\
\label{eq:LFFock}
\eea
The light-front, 
Fock-space operator 
$a^\dagger_{k_i}$ 
creates an on-shell constituent, 
$a^\dagger_{k_i}  | 0 \rangle = |k_i^+, \bm{k}_{i \perp} \rangle$. 
The light-front wavefunction 
$\psi_{LF}(k_1, k_2, P)$
is symmetric under interchange of the constituent's momenta, 
and by virtue of the momentum conserving delta-function
always appears in the form 
$\psi_{LF}(k_1, P - k_1,P)$. 
We shall use  schematic  notation and write this simply as
$\psi_{LF}(k_1, P)$, 
or even
$\psi_{LF} (x_1, \bm{k}_{1\perp})$
in the hadron's rest frame, 
where  
$\bm{P}_\perp = \bm{0}$,  
with
$x_1 = k^+_1 / P^+$. 
While there are higher Fock-state contributions to the covariant bound-state wave function, 
we use a two-particle truncation throughout. 
The factor 
$\mathcal{Q}$ 
appearing in the Fock-space decomposition
is the charge, 
which enters the normalization condition 
\be \label{eq:LFnorm}
\mathcal{Q} 
=
\frac{1}{(2\pi)^3} 
\int \frac{dx \, d\bm{k}_\perp}{2 x(1-x)}
| \psi_{LF} (x, \bm{k}_\perp) |^2
.\ee
Using the number density operator, 
the natural choice for the total charge is 
$\mathcal{Q} = 2$.

Using light-front quantized fields, 
we can derive an expression for 
$I(k_1, k_2, P)$
using the Fock-space expansion of Eq.~\eqref{ftpsi}. 
This yields
\bea
I(k_1, k_2, P) 
=
(2 \pi)^3 \delta^{(+, \perp)} (P - k_1 - k_2)
\frac{2 P^+}{2 k_1^+ 2 k_2^+}
\psi_{LF} (k_1, P)
.\eea
Comparing with Eq.~\eqref{firsti}, 
we find
\bea
\psi_{LF}(k,P)
=
{k^+ (P^+ - k^+) \over\pi P^+}
\int_{-\infty}^\infty dk^-\Psi(k, P)
\label{lfwf0}
.\eea
The factors involving plus-components of momentum arise from treating the phase-space covariantly in the Fock-state expansion.

By contrast, 
the bound state
$| \bm{P} \rangle $
in the instant-time formulation is chosen to satisfy the covariant normalization, 
$\langle \bm{P}' | \bm{P}  \rangle = 2 P^0 (2\pi)^3 \delta ( \bm{P}' - \bm{P})$, 
and has the Fock-space expansion
\bea 
| \bm{P}  \rangle
=
\frac{1}{\sqrt{2 \mathcal{Q}}}
\int 
\frac{d \bm{k}_1}{2 E_{\bm{k}_1} (2\pi)^3}  
\frac{d \bm{k}_2}{2 E_{\bm{k}_2} (2\pi)^3}  
\;
\psi_{IF} (k_1, k_2, P) 
\;
2 P^0 (2\pi)^3 \delta ( \bm{P} - \bm{k}_1 - \bm{k}_2)
\;
a_{\bm{k}_1}^\dagger a_{ \bm{k}_2}^\dagger | 0 \rangle
.
\notag \\
\label{eq:IFFock}
\eea
The instant-form, Fock-space operator 
$a_{\bm{k}_i}^\dagger$
creates an on-shell constituent
$a_{\bm{k}_i}^\dagger = | \bm{k}_i \rangle$.
Although we use a similar notation for Fock-space operators
in the instant and light-front forms, 
they are not related by a finite Lorentz transformation 
(only by a boost to infinite momentum). 
The instant-form wave function,
$\psi_{IF}(k_1, k_2, P)$,
is symmetric under interchange of the constituent's momenta, 
and by virtue of the momentum conserving delta-function
always appears in the form 
$\psi_{IF}(k_1, P - k_1,P)$. 
We shall use schematic  notation and write this simply as
$\psi_{IF}(\bm{k}_1, \bm{P})$, 
or 
$\psi_{IF}(\bm{k}_1)$ in the hadron's rest frame, 
$\bm{P} = \bm{0}$. 
The total charge 
$\mathcal{Q}$ 
enforces the rest-frame normalization condition 
\be \label{eq:IFnorm}
\mathcal{Q} 
=
\frac{1}{(2\pi)^3} 
\int \frac{d\bm{k}}{2  E_{\bm{k}}^2} \,
| \psi_{IF} (\bm{k}) |^2
.\ee
In general,
the Fock-state expansion is  expected to be much more complicated in the instant form
because of the need to deal with vacuum fluctuations.

In the instant form of dynamics, 
the energy and Lorentz boosts are dynamical operators, 
and the initial conditions are sepcified on the boundary 
$x^0=0$. 
Thus we  define an instant form version, 
$I^0(k_1, k_2, P)$, 
of the integral 
$I(k_1, k_2, P)$:
\bea 
I^0(k_1, k_2, P)
\equiv 
\int d^4x_1d^4x_2 \,
\delta( x_1^0)
\delta(x_2^0) \
\Psi(x_1,x_2,P)
\exp(ik_1\cdot x_1+ik_2 \cdot x_2)
.\eea
This integral produces a projection of the covariant Bethe-Salpeter wave function analogous to that  in Eq.~\eqref{firsti}. 
Using the instant-form Fock state expansion Eq.~\eqref{eq:IFFock},
the instant-form wavefunction 
$\psi_{IF}(\bfk, \bm{P})$ 
is given by
\bea 
\psi_{IF}{(\bfk,\bm{P})}
=
{E_{\bm{k}} \, E_{\bm{P} - \bm{k}} \over\pi P^0}
\int_{-\infty}^\infty\Psi(k,P) 
\, dk^0
.\label{ifwf0}
\eea
Our aim is to elucidate the differences and connections between 
$\psi_{LF}$ 
and 
$\psi_{IF}$.

\section{Toy Model}
\label{toy}

\begin{figure}
\begin{center}
\epsfig{file=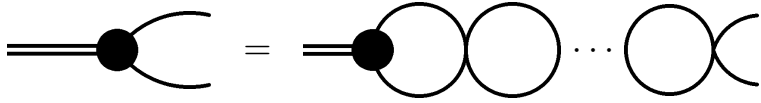,width=10cm}
\end{center}
\caption[Bethe-Salpeter equation for a point interaction.]{Bethe-Salpeter equation for a point interaction. The state is bound by the infinite chain of bubbles.}
\label{f:toy}
\end{figure}

Above we have discussed the covariant BSE for two-body bound states.  
In this section, 
we consider a toy model for the BSE that is exactly soluble. 
The solution will enable us to compare and contrast instant-form dynamics and light-front dynamics all while maintaining exact covariance.

One can obtain the simplest soluble BSE by choosing a point-like interaction for the kernel
$K(k,k';P)$ 
in Eq.~\eqref{bs}, 
namely 
$K(k,k';P) = g$, 
where 
$g$ 
is a coupling constant. 
The two scalar particles that make up the scalar bound state thus interact infinitely many times according to the BSE to bind the state. 
For the point-like interaction, 
a bubble chain is generated by the BSE, 
and is shown in Figure \ref{f:toy}. 
With this choice of interaction, 
the bound state equation simplifies tremendously. 
Since the kernel is independent of momentum, 
the only 
$k'$-dependence that remains in Eq.~\eqref{bs} is in 
$\Psi(k',P)$, 
and this quantity is subsequently integrated over all 
$k'$. 
The integration merely produces a constant  that can be absorbed into the overall normalization of the wavefunction.
Thus we are left with the solution
\begin{equation} \label{eqn:soln}
\Psi(k,P) = i g\; G(k, P-k)
,\end{equation}
where a proportionality constant is set to unity. 
The Bethe-Salpeter equation 
for the vertex
$\Gamma(k,P)$ 
also determines the mass, 
$M^2 = P^2$,
of the bound state via the consistency equation
\begin{equation}
1 = i g \int \frac{d^4 k}{(2\pi)^4} G(k,P -k) 
.\end{equation}
For simplicity, 
we do not discuss the necessary regularization, 
and treat the coupling 
$g$ 
as a renormalized parameter.

The single-particle propagator has the basic Klein-Gordon form, 
so the two-particle disconnected propagator 
is a product of these Klein-Gordon propagators. 
By virtue of Eq.~\eqref{eqn:soln}, 
the covariant Bethe-Salpeter wavefunction is
\begin{equation}
\Psi(k,P) = -ig [k^2 - m^2 + i \varepsilon]^{-1} [(P-k)^2 - m^2 + i
\varepsilon]^{-1} \label{bswf}
.\end{equation}
Here we have labeled the constituent mass by $m$.
This is a four-dimensional analogue of the usual Schr\"odinger wave function. 
There is, however, an important distinction. 
We also know the time dependence of the wave function---the time evolution governed by the Hamiltonian operator
is automatically included because of the necessity of covariance. 
Moreover, 
we know from the Poincar\'e algebra that there are other dynamical operators besides the energy.
As to which operators are kinematical depends upon the form of dynamics chosen.

\subsection{Rest-frame wave functions}       %

We shall next compute the instant-form wave function using \eq{ifwf0}
as evaluated in the rest frame. 
Given our solution to the BSE, 
Eq.~\eqref{bswf}, 
we can carry out this projection onto the initial surface.
The integration can be done using the residue theorem bearing in mind the four poles of the integrand:
$k^0 = \pm E_{\bm{k}} \mp i \varepsilon$, 
and
$M \pm E_{\bm{k}} \mp i \varepsilon$.
We find
\begin{equation} \label{eqn:rest}
\psi_{IF}(\bm{k},\bm{0}) = -\frac{2 g}{M} \,  \frac{\sqrt{\bm{k}^2 + m^2}}{M^2 - 4 (\bm{k}^2 + m^2)} .
\end{equation}
Notice the wavefunction is manifestly rotationally invariant. 
This is indicative of the kinematic nature of the generators of rotations in the instant form.

In the front form of dynamics, one is interested in the properties of physical states along the advance of a wavefront of light.
The objects of front-form dynamics are the light-cone wave functions which are projections onto the initial surface 
$x^+=0$.
In analogy with the instant form, one refers to 
$x^+$ 
as light-cone time, 
and its Fourier conjugate 
$k^-$ 
as light-front energy. 
In the front form, 
the energy is a dynamical operator along with two rotation operators corresponding to two independent 
rotations of the wavefront of light.  
In contrast with the instant form, light-front Lorentz boosts are kinematical. 
We use  Eq.~\eqref{lfwf0}, 
and work in the hadronic rest-frame, 
$\bm{P}_\perp = \bm{0}$, 
to define
$\psi_{LF}(x,\bfk_\perp)$,
with 
$x=  k_1^+ / P^+ = k^+ /  P^+$.
The light-cone wavefunction  corresponding to \eq{bswf} 
is found by contour integration of \eq{lfwf0}  to be 
\begin{equation}
\psi_{LF}(x, \mathbf{k}_\perp) = -  g  
\frac{ \theta[x(1-x)] }{M^2 - \frac{\bm{k}_\perp^2+m^2}{x(1-x)}}
\label{lfgood}\ee
Note that the full rotational symmetry of the rest-frame wavefunction is not manifest.

We now inquire as to how the 
$IF$ 
and 
$LF$ 
wave functions are related to each other. 
In the literature, 
the rest frame 
$IF$ 
wave function is converted into the rest frame the light-cone wave function 
by introducing an auxiliary variable,  
$x$, 
using Eq.~\eqref{eqn:bad}.
This variable has a physical interpretation as the fractional 
plus-component of momentum in the center of mass system 
of two free particles. 
Inverted this relation between 
$x$ 
and 
$k^3$ reads~%
\cite{Tiburzi:2000je}
\begin{equation} \label{eqn:variable}
k^3 = \left( x - \frac{1}{2} \right) \sqrt{\frac{\bm{k}_\perp^2 + m^2}{x(1-x)}}
.\end{equation} 
Simple algebra yields the relation
\begin{equation} 
4 ( \bm{k}^2 + m^2) ={\bfk_\perp^2+m^2\over x(1-x)}
\label{invert}
,\end{equation}
from which we deduce
\begin{equation}
\psi_{IF}(\bfk,0)\rightarrow\psi_{IF}(x,\bfk_\perp) =
- {g \over M}\sqrt{\bfk_\perp^2+m^2\over x(1-x)} \, \frac{1}{M^2 - {\bfk_\perp^2+m^2\over x(1-x)}}
.\label{ifwf}\end{equation}
This bears a resemblance to the front-form wavefunction in the rest
frame, Eq.~\eqref{lfgood}, but the instant-form wave function carries
an additional factor of 
$E_{\bm{k}} /M $. 
This is a clear and major difference. 
One cannot interpolate between the instant form and light-front form of the wave function.

One suspects that the two forms become equivalent in the non-relativistic limit. 
This limit is defined by replacing 
$\sqrt{\bm{k}^2+m^2}$
with 
$m$, 
so that \eq{eqn:bad} becomes
\bea
x\rightarrow \frac{1}{2}+ \frac{k^3}{2 m} 
\label{nonrel}
.\eea 
In the non-relativistic limit, 
we write the bound-state mass in terms of the constituent masses and a small binding energy 
$B  > 0$, 
namely
$M=2m-B$. 
Expanding about 
$B =0$ 
to linear order, 
and replacing the factors
$E_\bfk$ 
that appear in the relativistic phase space by 
$m$,
\eq{ifwf} then becomes
\bea\label{fake}
\psi_{IF}(x,\bfk_\perp)
\rightarrow 
-g\frac{\theta[x(1-x)]}{M^2 - \frac{\mathbf{k}_\perp^2+m^2}{x(1-x)}},
\eea
the same as 
\eq{lfgood}. 
The 
$\theta$-function 
appears as a result of \eq{invert}.
We see that the wave functions of the two forms become identical only in the non-relativistic limit.
But there is no reason to suspect that this limit should be valid
because the wave functions fall off very slowly in momentum space.
The only way to tell is to look at specific matrix elements.

It has been convenient to examine electromagnetic form factors. 
Truncating at the lowest Fock state, 
the expression for the electromagnetic form factor in terms of the 
front-form wave function is given by~%
\cite{Drell:1969km,West:1970av}
\bea\label{lfff}
F_{LF}(Q^2)={1\over (2\pi)^3}\int \psi_{LF}(x,\bfk_\perp)\psi^*_{LF}(x,\bfk_\perp+(1-x)\bfQ_\perp){ dx \, d\bm{k}_\perp \over2x(1-x)}
,\eea
where the momentum transfer appears as  
$q^2 =  - Q^2 = - \bm{Q}_\perp^2$,
in a frame where 
$q^+ = 0$.
A virtue of the light-front formulation is that the boost required between initial and final states in Eq.~\eqref{lfff} is kinematical. 
The instant-form expression also requires a boosted wave function, 
however, 
instant-form boosts are dynamical. 
This complicates the interpretation of the form factor in terms of instant-form quanta. 
For example, 
it is well-known that boosting does not conserve particle number. 
With initial and final states differing in particle number, 
the instant-form form factor consequently cannot be the Fourier transform of a charge density. 
On the other hand, 
due to the kinematic nature of light-front boosts, 
the form factor has an interpretation in terms of the transverse charge density of quanta in the infinite momentum frame~%
\cite{Soper:1976jc,Burkardt:2000za,Burkardt:2002hr,Miller:2007uy,Carlson:2007xd,Miller:2009qu}.

For our toy model ($TM$), 
we use \eq{lfgood} in the above expression to find
\bea\label{lftm}
F_{LF}^{TM}(Q^2)={g^2\over (2\pi)^3}\int {1\over M^2-{\bfk_\perp^2+m^2\over x(1-x)}} {1\over M^2-{[\bfk_\perp+(1-x)\bfQ_\perp]^2+m^2\over x(1-x)}} 
{ dx \, d \bm{k}_\perp \over2x(1-x)}
\eea
On the other hand, the use of the ersatz light-front wave function \eq{ifwf} in \eq{lfff} would lead the appearance of a factor 
\be
\frac{1}{x(1-x)}
\sqrt{
(\bfk_\perp^2+m^2)
([\bfk_\perp+(1-x)\bfQ_\perp]^2 +m^2)}
\notag
\ee 
in the integrand of \eq{lftm}. 
This would lead to divergences in the integrals over both
$x$,
and
$d \bm{k}_\perp$.
The form factor of this toy model was studied extensively for several different situations 
in
Ref.~%
\cite{Miller:2009sg}. 
There, 
it was shown that the equal-time wave function in the
rest frame has no direct connection with the form factor,
but the exact covariant evaluation of the form factor is indeed obtained using the expression \eq{lftm}. 
In the
non-relativistic limit, 
the light-front and equal-time form factors do coalesce to the same result. 
However, 
this limit is satisfied for very limited kinematics, 
$B/M <0.002$. 
Thus the correspondence embodied by using the simple expression \eq{eqn:bad}
does not work for the simplest possible toy model.

An additional ingredient common to the popular recipe for making a light-front wave function
involves including a Jacobian factor in order to preserve the normalization of the wave function. 
The normalization of the 
ETRF 
wave function in Eq.~\eqref{eq:IFnorm}
will pick up a Jacobian,
$J = \partial k^3 / \partial x$, 
if we view
Eq.~\eqref{eqn:variable}
as a change of variables. 
Taking into account the relativistic phase space factors, 
Eq.~\eqref{eq:IFnorm}
will have exactly the form of  
Eq.~\eqref{eq:LFnorm}
provided we make the identification
\be
\psi_{JIF} (x, \bm{k}_\perp ) 
\equiv
\sqrt{M} 
\left[ 
\frac{\bm{k}_\perp^2 + m^2}{x(1-x)} 
\right]^{-1/4} 
\psi_{IF} (x, \bm{k}_\perp ) 
\longrightarrow 
\psi_{LF} (x, \bm{k}_\perp )
.\ee
For the toy model, 
however, 
the Jacobian modified instant-form wave function (JIF)
\be \label{eq:JIF}
\psi_{JIF} (x, \bm{k}_\perp ) 
=
- \frac{g}{\sqrt{M}}
\left[ \frac{\bm{k}_\perp^2 + m^2}{x(1-x)} \right]^{1/4}
\frac{1}{ M^2 - \frac{\bm{k}_\perp^2 + m^2}{x(1-x)}}
,\ee
is still not the light-front wave function
$\psi_{LF}(x, \bm{k}_\perp)$ in Eq.~\eqref{lfgood}. 
A factor of the Jacobian squared, 
$J^2$, 
will produce the light-front wave function in this model, 
however, 
there is no justification to include two powers of the Jacobian.

To properly derive the instant-form expression for the form factor in the toy model, 
one starts from the covariant triangle diagram, 
and performs the projection onto equal-time by integrating over the loop energy, 
$k^0$. 
The time-ordered diagrams that result, 
see for example~%
\cite{Sawicki:1991sr}, 
contain non-wave function terms.  
The presence of such terms demonstrates that the form factor in the instant-form dynamics
cannot be related to the Fourier transform of a charge density. 
In the toy model, 
the instant-form boost leads to non-trival effects, 
which nonetheless can be determined explicitly. 
In QCD, 
in contradistinction, 
the boost is too complicated to allow a general solution,
although there has been progress for small momentum~%
\cite{Rocha:2009xq}.

\subsection{Boosting to the infinite momentum frame}

\def\bfP{{\bf P}}
The  only  way to relate the 
$IF$ 
and 
$LF$ 
wave functions is by boosting the 
$IF$ 
wave function  to the infinite momentum frame. 
In that frame,
it becomes the same as the 
$LF$ 
wave function~%
\cite{Tiburzi:2004ye}.
The way to see this is to obtain the 
$IF$ 
wave function in a frame in which the 3-component of the momentum takes on an arbitrary value, 
and then let this value to approach infinity.
To do this,  
we must first re-evaluate the expression \eq{ifwf0} in a frame in which the system is moving with momentum 
$\bfP$ 
in a direction associated with the 3-axis. 
With the bound state energy 
$P^0$ 
given by 
$P^0 = \sqrt{\bm{P}^2 + M^2}$, 
evaluation of the contour integration of \eq{ifwf0} using the toy model wave function 
$\Psi(k,P)$ in~\eq{bswf} yields the wave function:
\bea
\psi_{IF}(\bfk,\bfP)
=
- \frac{g}{2 P^0}
\left[{1\over P^0 -  E_{\bm{k}} - E_{\bm{P} - \bm{k}}} - {1\over P^0 + E_{\bm{k}}+ E_{\bm{P} -\bm{k}}} \right]
\label{topt}
\eea
The first term in \eq{topt} corresponds to a time-ordered graph with particle propagation, 
while the second term corresponds to particles propagating backwards in time.

We wish to take the limit of $P\rightarrow\infty$. 
To this end,
define the third component of 
$\bfk$
to be 
$x P$, 
so that the third component of of 
$\bfP-\bfk$ 
is 
$(1-x)P$.
In the limit that 
$|\bfP|$ 
approaches infinity, 
the wave function of \eq{topt} vanishes unless 
$0<x<1$. 
In that case,
the following limits hold 
\bea \lim_{P\rightarrow\infty} E_{\bm{k}} \label{lim1}
&=& xP +{\bfk_\perp^2+m^2\over 2 x P},\\
\lim_{P\rightarrow\infty} E_{\bm{P} -\bm{k}}
&=&
(1-x)P +{\bfk_\perp^2+m^2\over 2 (1-x) P},\\
\lim_{P\rightarrow\infty} P^0
&=&
P +{M^2\over 2P}
\label{lim3}
.\eea 
For large values of $P$, 
only the first (or wave function)  
term of \eq{topt} is non-vanishing. 
Taking the limit of \eq{topt} as $P$ approaches infinity leads immediately to the result
\bea \label{done}
\lim_{P\rightarrow\infty}\psi_{IF}(\bfk,\bfP)=\psi_{LF}(x,\bfk_\perp)
.\eea
While we have demonstrated this result using the toy model wavefunction, 
we remark that the instant-form Fock space expansion in Eq.~\eqref{eq:IFFock} 
can be boosted to infinite momentum.
One arrives at Eq.~\eqref{eq:LFFock} which demonstrates the equivalence in Eq.~\eqref{done} more generally.

\section{Other Models}

We study more elaborate models defined by  
interactions other than point-like coupling, using the formalism of \cite{Carbonell:2008tz}.
In the BSE, 
the interaction kernel 
$K$ 
is given by irreducible Feynman diagrams. 
Using any finite set of them is an approximation to the theory under consideration.
If the kernal is given by a set of
Feynman graphs~%
\cite{Nakanishi:1969ph,Nakanishi:1988hp}, 
the Minkowski space BS amplitude~Eq.~\eqref{bs} is
found in terms of the Nakanishi integral representation~%
\cite{Nakanishi:1963zz}:
\begin{eqnarray}\label{bsint}
\Phi(k;P)&=&-{i\over \sqrt{4\pi}}\int_{-1}^1dz\int_0^{\infty}d\gamma
\frac{g(\gamma,z)}{\left[\gamma+m^2
-\frac{1}{4}M^2-k^2-P\cdot k\; z-i\varepsilon\right]^3}.
\end{eqnarray}
The weight function 
$g(\gamma,z)$ 
itself is not singular, 
whereas the singularities of the BS amplitude are fully reproduced by this integral. 
For example, 
if one sets 
$g(\gamma,z)= \sqrt{ 4 \pi} \,  g$ 
and calculates the integral, the result is 
the product of two free propagators appearing in \eq{bswf}.

The wave function in the ETRF is obtained by using \eq{bsint} in \eq{ifwf0},
with the result
\begin{eqnarray}
\psi_{IF}(\bfk,\mathbf{0})
&=&
-
{1 \over \sqrt{4\pi}}
{ 3(\bfk^2+m^2 ) \over 8 M} 
\int_{-1}^1dz\int_0^{\infty}d\gamma 
\frac{g(\gamma,z)}{\left[\gamma+ \bfk^2 + m^2
-\frac{1}{4}M^2(1-z^2) \right]^{5/2}}.
\label{phiinst}\end{eqnarray}
The light-front wave function 
$\psi_{LF}(\bfk_\perp,x)$ 
is defined as before by an integration over 
$k^-$ 
as in \eq{lfwf0}. 
Substituting \eq{bsint} into \eq{lfwf0},
the two-body light-front wave function is found to be \cite{Carbonell:2008tz}:
\begin{equation}\label{bs10}
\psi_{LF}(\bfk_{\perp},x)
=-\frac{1}{\sqrt{4\pi}}\int_0^{\infty}\frac{x(1-x) \, g(\gamma,1-2x)\, d\gamma}
{\left[ \gamma+\bfk_{\perp}^2+m^2-x(1-x)M^2 \right]^2}\ .
\end{equation}

Our next task is to compare the expressions in \eq{phiinst} and \eq{bs10}. 
It is possible to show in general that:
the non-relativistic ($NR$) limit of these equations is the same, 
and 
boosting the ETRF wave function to the infinite momentum frame results in the light-front wave function. 
We handle this former first. 
Using the replacement \eq{nonrel} in the light-front wave function \eq{bs10}, 
and keeping terms linear in the binding energy, 
one obtains
\begin{equation}\label{bs11}
\psi_{LF}^{NR}(\bfk)
=-\frac{1}{4\sqrt{4\pi}}\int_0^{\infty}\frac{g(\gamma,0) \, d\gamma}
{\big(\gamma+\bfk^2+m^2- \frac{1}{4} M^2 \big)^2}\ .
\end{equation}
Next work with the instant-form wave function, \eq{phiinst}. 
The mass-squared, 
$M^2\approx 4m^2-4mB$, 
is a  large quantity in the non-relativistic limit. 
Thus we may use 
$g(\gamma,z)\approx g(\gamma,0)$, 
so that the integral over 
$z$ 
can be performed. 
Note also that energies appearing in phase-space factors are replaced by constituent masses in the $NR$ limit. 
Then we have
\begin{eqnarray}
\psi_{IF}^{NR}(\bfk,\mathbf{0})
&=&
- {1\over \sqrt{4\pi}} {m^2 \over  M} 
\int_0^{\infty} 
\frac{\left[M^2+6(\gamma+ \bfk^2 + m^2- \frac{1}{4} M^2)\right]}
{\left[M^2+4(\gamma+ \bfk^2 + m^2- \frac{1}{4} M^2 \right]^{3/2}}
\frac{g(\gamma,0) \, d\gamma}{(\gamma +\bfk^2 +m^2- \frac{1}{4} M^2)^2}
.\notag \\
\label{phiinstnr}\end{eqnarray}
The ratio of bracketed terms in the above expression reduces to 
$1/(2m)$ 
in the $NR$ limit. 
In that case,  
the results \eq{phiinstnr} and \eq{bs11} become identical. 
Thus in general, 
the correspondence between the instant-form and front-form wave functions is obtained when the non-relativistic limit is valid. 
This is expected because in the non-relativistic limit the wave functions are frame-independent.

To demonstrate the equivalence of the light-front wave function and the equal-time wave function in the infinite momentum frame, 
we return to Eq.~\eqref{bsint} to derive the equal-time wave function in an arbitrary frame. 
We find
\bea 
\psi_{IF}(\bm{k}, \bm{P})
&=&
- \frac{1}{\sqrt{4 \pi}}
\frac{3 E_{\bm{k}} \, E_{\bm{P} - \bm{k}}}{8 P^0}
\int_{-1}^{1} dz 
\int_0^\infty d\gamma
\; g(\gamma,z)
\notag \\
& & 
\times
\Big[ 
\gamma 
+
\bm{k}^2
- 
(1 - z) \bm{k} \cdot \bm{P}
+
\frac{1}{4} ( 1 - z)^2 \bm{P}^2
+ 
m^2
-
\frac{1}{4} (1 - z)^2 M^2 
\Big]^{-5/2}
.\eea
Using the limits in Eqs.~\eqref{lim1}--\eqref{lim3}, 
the wave function vanishes as 
$1 / P^{4}$
when 
$P \to \infty$.
This is true for all values of 
$z$, 
except in the region around
$z = 1 - 2 x$. 
To obtain the non-vanishing contribution in the infinite momentum frame, 
we must thus replace 
$g(\gamma, z) = g ( \gamma, 1 - 2 x)$.
This replacement enables us to perform the 
$z$-integration explicitly, 
and subsequently take the 
$P \to \infty$ 
limit. 
This procedure yields the equivalence
\be
\lim_{P \to \infty}
\psi_{IF} ( \bm{k}, \bm{P} ) 
= 
\psi_{LF} ( x, \bm{k}_\perp )
,\ee
for any wave function for which the Nakanishi integral representation Eq.~\eqref{bsint} is valid. 
To compare the ETRF wave function to the light-front wave function using 
the recipe in Eq.~\eqref{eqn:bad}, 
however,
we need to know about the functional form of 
$g(\gamma,z)$. 
This is most easily done using specific models, to which we now turn.

\subsection{Rotationally invariant light-front model}                      %

To investigate further the relation between the wave functions in Eqs.~\eqref{phiinst} and \eqref{bs10}, 
we adopt a model. 
We may enforce rotational invariance 
$RI$
in the light-front wave function by choosing 
$g(\gamma,z)$ 
to have a particular form:
\bea
g^{RI}(\gamma,z)
=
4 g_0 \, \delta(\gamma) (1-z^2), 
\label{gri}
\eea
where 
$g_0$ 
is a constant. 
Using \eq{gri} in \eq{bs10} 
leads to the light-front wave function
\bea
\psi^{RI}_{LF}(\bfk_{\perp},x)
=
- {g_0\over\sqrt{4\pi}}
\frac{16}{\left[ M^2 - \frac{\bm{k}_\perp^2 + m^2}{x (1-x)} \right]^2}
.\eea
With the help of the variable 
$\kappa$, 
defined by 
\begin{equation}
\kappa^2 = m^2 - \frac{1}{4} M^2
,\end{equation}
we can cast the light-front wave function into a suggestive form. 
Using the inverse of the recipe, 
Eq.~\eqref{eqn:variable}, 
we can introduce the variable 
$k^3$ 
to make the light-front wave function appear as a rotationally invariant
instant-form wave function
\bea
\psi^{RI}_{LF}(\bfk_{\perp},x)
\to
- {g_0\over\sqrt{4\pi}}
{1\over \left(\bfk^2+\kappa^2\right)^2}
.\label{lri}\eea
This wave function has the same form as that for the lowest $s$-state of a hydrogenic atom.

The corresponding rest-frame, instant-form wave function is obtained by
using \eq{gri} in \eq{phiinst}
\bea
\psi^{RI}_{IF}(\bfk)
&=&
- \frac{g_0}{ \sqrt{4 \pi}} \,
{2 \over M}
\frac{\sqrt{\bm{k}^2 + m^2}}
{ \left(\bm{k}^2+\kappa^2 \right)^2}
\label{iri}
.\eea
In this case, one can compare the two forms \eq{iri} and \eq{lri} having already used \eq{eqn:bad}. 
It is readily apparent that the two forms are very different. 
For example for large values of 
$\bfk^2$, 
the former falls as 
$1/ | \bm{k}|^3$, 
while the latter falls as 
$1/ \bm{k}^4$. 
Once again, 
we see that  the relation between the  
rest-frame wave function and the light-front wave function cannot be seen using a simple transformation.

As with the toy model, 
including the Jacobian factor in converting the instant-form wave function, 
as in Eq.~\eqref{eq:JIF}, 
does not produce the light-front wave function. 
The ratio of the Jacobian modified instant-form wave function
to the true light-front wave function is not unity, 
\be \label{eq:ratio}
\frac{
\psi^{RI}_{JIF}(x, \bm{k}_\perp ) 
}{
\psi^{RI}_{LF} (x, \bm{k}_\perp)
}
=
\frac{1}{\sqrt{M}}
\left[
\frac{\bm{k}_\perp^2 + m^2}{x(1-x)}
\right]^{1/4}
.\ee
Curiously enough, 
this ratio, 
while not unity,
is the same in the $RI$ model as in the toy model of Sect.~\ref{toy}. 
This coincidence owes to the simplicity of the models considered, 
however,
not an underlying principle, 
as the final example demonstrates.

\subsection{Wick-Cutkosky (WC) Model} 

Let us consider a field theoretic example. 
Exact solutions to the Bethe-Salpeter equation in the ladder approximation are known. 
In the WC model~%
\cite{Wick:1954eu,Cutkosky:1954ru}, 
two scalars are bound by scalar exchange, 
and the function 
$g(\gamma,z)$ 
has the form
\bea
g^{WC}(\gamma,z)
=
\delta(\gamma) \lambda \, (1-|z|)
,\label{gwc}\eea
with the constant 
$\lambda$
defined in terms of parameters of the model,
$\lambda = 2^6 \pi \sqrt{m} \, \kappa^{5/2}$. 
Given this form for 
$g(\gamma, z)$,
we evaluate the instant and light-front 
wave functions by using \eq{gwc} in \eq{phiinst} and \eq{bs10}. 
We find the instant-form wave function to be:
\bea\label{wcinst}
\psi^{WC}_{IF}(\bfk)
=
-{\lambda \over  \sqrt{4\pi} \, M^3}
\frac{\sqrt{\bfk^2+m^2}}{(\bm{k}^2 + \kappa^2)^2}
\left[
\bm{k}^2 + \kappa^2 + \frac{1}{2} M^2 -  \sqrt{ (\bfk^2+m^2 )(\bm{k}^2 + \kappa^2) } 
\,
\right]
.\eea
In the non-relativistic limit, 
this wave function becomes identical to that of the ground-state hydrogenic atom.
Away from this limit, 
the wave function contains relativistic phase-space factors, 
and the effects of retardation.   
In the asymptotic limit, 
the wave function has the behavior
\bea
\lim_{| \bm{k} | \rightarrow \infty} 
\psi^{WC}_{IF}(\bfk)
=
{3 \lambda \over 8 \sqrt{4\pi} \, M} \, 
\frac{1}{ |\bm{k}|^3}
.\eea
We find the light-front wave function to be given by
\bea\label{lfwc}
\psi^{WC}_{LF}(\bfk_{\perp},x)
=
-
\frac{\lambda}{\sqrt{4\pi}} 
\,
{1 - | 1 - 2 x| \over x(1-x)}
\,
\frac{1}
{\left[ M^2 - \frac{\bfk_{\perp}^2+m^2}{x(1-x)} \right]^2}
.\eea

Immediate inspection indicates that the wave functions of \eq{wcinst} and \eq{lfwc} are very different.
The light-front wave function falls off faster than the instant-form wave function at large transverse momentum. 
We can try to relate the two wave functions by using the relation in \eq{invert}. 
The ratio of the transformed instant-form wave function to the light-front wave function is considerably different
than unity
\bea
\frac{\psi^{WC}_{IF}(x, \bm{k}_\perp) }{ \psi^{WC}_{LF}( x, \bm{k}_\perp )}
&=&
\frac{2}{M^3} \frac{x (1-x)}{1 - | 1 - 2 x|}
\sqrt{\frac{\bfk_{\perp}^2+m^2}{x(1-x)}}
\notag \\
&&
\times
\left(
\frac{\bfk_{\perp}^2+m^2}{x(1-x)} + M^2 - \sqrt{\frac{\bfk_{\perp}^2+m^2}{x(1-x)} \left[ \frac{\bfk_{\perp}^2+m^2}{x(1-x)} - M^2 \right]}
\,
\right)
.\eea
A simple substitution as given by \eq{eqn:bad} cannot relate the instant and light-front wave functions.
Including the Jacobian factor via Eq.~\eqref{eq:JIF} 
does not simplify the ratio
$\psi^{WC}_{JIF}(x, \bm{k}_\perp) / \psi^{WC}_{LF}( x, \bm{k}_\perp )$. 
This ratio, 
moreover, 
is considerably different than the common value, 
Eq.~\eqref{eq:ratio}, 
found in the two simpler toy models.

\section{Summary}

We use simple covariant models for which the solutions of the Bethe-Salpeter equation can be obtained. 
This  allows us to explore both the instant and front-form wave functions. 
The structure of these wavefunctions is related to the respective kinematic subgroups of the Poincar\'e algebra. 
Moreover, 
a fully covariant starting point allowed us a simple way to correctly formulate three-dimensional dynamics. 
We find that it is not possible to use the simple transformation \eq{eqn:bad} 
to relate the rest-frame instant-form wave function with the light-front wave function. 
The only known way to do this is to boost the rest-frame instant-form wave function to the infinite momentum frame.

There is an interesting relation between integrals of $IF$ and $LF$ wave functions that can be derived,
similar relations have been suggested in~\cite{Broniowski:2007si,Miller:2007uy}. 
The projection onto the space-time point
$x^0 = x^3 = 0$
is a unique place where the 
$IF$ 
wave function can be related to the 
$LF$ 
wave function. 
This is because at this point we also have
$x^+ = x^- = 0$, 
so that equal time also corresponds to equal light-front time.  
Consider the bound state in an arbitrary frame
with 
$P^\mu = ( \sqrt{\bm{P}^2 + M^2}, \bm{P} )$. 
Integrating the 
$IF$
wave function over the third-component of momentum
projects onto 
$x^3 = 0$.
Carrying out this projection, 
we find
\bea
\int_{-\infty}^\infty dk^3 
\frac{P^0}{E_{\bm{k}} E_{\bm{P} - \bm{k}} }
\, 
\psi_{IF}(\bm{k}, \bm{P}) 
&=&
\frac{1}{\pi} \int_{-\infty}^\infty 
dk^0 dk^3 \,
\Psi(k, P)
\notag \\
&=&
\frac{1}{2 \pi} 
\int_{-\infty}^\infty 
dk^- dk^+ \,
\Psi(k, P)
\notag \\
&=&
\int_0^1 \frac{dx}{x (1-x)} 
\psi_{LF}( x, \bm{k}_\perp - x \bm{P}_\perp)
\label{convention}
,\eea
which shows that integrals over the 
$IF$ 
and 
$LF$ 
wave functions are identical. 
This relation also elucidates why the 
$IF$ 
and 
$LF$
wave functions vanish with different powers of 
$|\bm{k}_\perp|$.

In the rest frame,
$\bm{P} = \bm{0}$,
one can derive a relation between the impact-parameter dependent 
$LF$ 
wave function, 
$\psi_{LF} (x, \bm{b}_\perp)$, 
defined by
\be
\psi_{LF} (x, \bm{b}_\perp)
=
\int \frac{d \bm{k}_\perp}{( 2 \pi)^2 }  \, e^{i \bm{b}_\perp \cdot \bm{k}_\perp } \,
\psi_{LF} (x, \bm{k}_\perp)
.\ee
From Eq.~\eqref{convention}, 
we find 
\bea
\int_{-\infty}^\infty 
dk^3
\int_{-\infty}^\infty 
\frac{d \bm{k}_\perp} { ( 2\pi)^2}
\frac{ M }{ \bm{k}^2 + m^2 }
\, 
e^{i \bm{b}_\perp \cdot \bm{k}_\perp } 
\,
\psi_{IF}(\bm{k})
&=&
\int_0^1 \frac{dx}{x (1-x)} 
\psi_{LF}( x, \bm{b}_\perp)
\label{trans}
,\eea
which is similar to the transversity relation found in~%
\cite{Broniowski:2009dt}. 
As a consistency check, 
it is trivial to verify this identity using the Nakanishi integral representation of the
$IF$ and $LF$ wave functions.
Although there is no simple recipe to cook up a light-front wave function from an equal-time, 
rest-frame wave function,
Eqs.~\eqref{convention} and \eqref{trans}
provide rigorous relations between their integrals. 
Given the phenomenological utility of light-front wave functions, 
we intend to explore whether further such relations exist.

%
\begin{acknowledgments}
This work is supported by the U.S.~Department of Energy, 
under Grants
No.~DE-FG02-97ER-41014 (G.A.M.), 
and
No.~DE-FG02-93ER-40762 (B.C.T.). 
B.C.T.~acknowledges the partial support of the Institute for Nuclear Theory during a visit that initiated this project.
\end{acknowledgments}
%

\bibliography{relateBIB}

\end{document}